\documentclass[preprint, authoryear,
nopreprintline]{elsarticle} 

\usepackage[hyphens]{url}

\usepackage{graphicx}

\usepackage[T1]{fontenc}
\usepackage{lmodern}
\usepackage{amssymb,amsmath}
\usepackage{lineno} 
\usepackage{ifxetex,ifluatex}
\usepackage{fixltx2e} 
\IfFileExists{upquote.sty}{\usepackage{upquote}}{}
\ifnum 0\ifxetex 1\fi\ifluatex 1\fi=0 
  \usepackage[utf8]{inputenc}
\else 
  \usepackage{fontspec}
  \ifxetex
    \usepackage{xltxtra,xunicode}
  \fi
  \defaultfontfeatures{Mapping=tex-text,Scale=MatchLowercase}
  
\fi
\IfFileExists{microtype.sty}{\usepackage{microtype}}{}
\usepackage[]{natbib}
\ifxetex
  \usepackage[setpagesize=false, 
              unicode=false, 
              xetex]{hyperref}
\else
  \usepackage[unicode=true]{hyperref}
\fi
\hypersetup{breaklinks=true,
            bookmarks=true,
            pdfauthor={},
            pdftitle={rcpptimer: Rcpp Tic-Toc Timer with OpenMP Support},
            colorlinks=false,
            urlcolor=blue,
            linkcolor=magenta,
            pdfborder={0 0 0}}

\usepackage{color}
\usepackage{fancyvrb}

\DefineVerbatimEnvironment{Highlighting}{Verbatim}{commandchars=\\\{\}}
\usepackage{framed}
\definecolor{shadecolor}{RGB}{248,248,248}
\newenvironment{Shaded}{\begin{snugshade}}{\end{snugshade}}

\newcommand{\AttributeTok}[1]{\textcolor[rgb]{0.13,0.29,0.53}{#1}}

\newcommand{\BuiltInTok}[1]{#1}

\newcommand{\CommentTok}[1]{\textcolor[rgb]{0.56,0.35,0.01}{\textit{#1}}}

\newcommand{\ControlFlowTok}[1]{\textcolor[rgb]{0.13,0.29,0.53}{\textbf{#1}}}
\newcommand{\DataTypeTok}[1]{\textcolor[rgb]{0.13,0.29,0.53}{#1}}
\newcommand{\DecValTok}[1]{\textcolor[rgb]{0.00,0.00,0.81}{#1}}

\newcommand{\FloatTok}[1]{\textcolor[rgb]{0.00,0.00,0.81}{#1}}
\newcommand{\FunctionTok}[1]{\textcolor[rgb]{0.13,0.29,0.53}{\textbf{#1}}}
\newcommand{\ImportTok}[1]{#1}

\newcommand{\KeywordTok}[1]{\textcolor[rgb]{0.13,0.29,0.53}{\textbf{#1}}}
\newcommand{\NormalTok}[1]{#1}
\newcommand{\OperatorTok}[1]{\textcolor[rgb]{0.81,0.36,0.00}{\textbf{#1}}}
\newcommand{\OtherTok}[1]{\textcolor[rgb]{0.56,0.35,0.01}{#1}}
\newcommand{\PreprocessorTok}[1]{\textcolor[rgb]{0.56,0.35,0.01}{\textit{#1}}}

\newcommand{\SpecialCharTok}[1]{\textcolor[rgb]{0.81,0.36,0.00}{\textbf{#1}}}

\newcommand{\StringTok}[1]{\textcolor[rgb]{0.31,0.60,0.02}{#1}}

\providecommand{\tightlist}{%
  \setlength{\itemsep}{0pt}\setlength{\parskip}{0pt}}

\usepackage{fvextra}
\DefineVerbatimEnvironment{Highlighting}{Verbatim}{
  breaksymbolleft={}, 
  showspaces = false,
  showtabs = false,
  breaklines,
  commandchars=\\\{\}
}

\begin{document}

\begin{frontmatter}

  \title{rcpptimer: Rcpp Tic-Toc Timer with OpenMP Support}
    \author[University of Duisburg-Essen]{Jonathan Berrisch%
  \corref{cor1}%
  }
   \ead{jonathan.berrisch@uni-due.de} 
      \affiliation[University of Duisburg-Essen]{
    organization={University of
Duisburg-Essen},city={Essen},country={Germany},}
    \cortext[cor1]{Corresponding author}
  
  \begin{abstract}
  Efficient code writing is both a critical and challenging task,
  especially with the growing demand for computationally intensive
  algorithms in statistical and machine-learning applications. Despite
  the availability of significant computational power today, the need
  for optimized algorithm implementations remains crucial. Many R users
  rely on Rcpp to write performant code in C++, but writing and
  benchmarking C++ code presents its own difficulties. While R's
  benchmarking tools are insufficient for measuring the execution times
  of C++ code segments, C++'s native profiling tools often come with a
  steep learning curve. The rcpptimer package bridges this gap by
  offering a simple and efficient solution for timing C++ code within
  the Rcpp ecosystem. This novel package introduces a user-friendly
  tic-toc class that supports overlapping and nested timers and OpenMP
  parallelism, providing nanosecond-level time resolution. Results,
  including summary statistics, are seamlessly passed back to R without
  requiring users to write any C++ code. This paper contextualizes the
  rcpptimer package within the broader ecosystem of R and C++ profiling
  tools, explains the motivation behind its development, and offers a
  comprehensive overview of its implementation. Supplementary to this
  paper, we provide multiple vignettes that thoroughly explain this
  package's usage.
  \end{abstract}
  
 \end{frontmatter}

\section{Introduction}\label{introduction}

The ever-increasing number of R packages reflects the popularity of R on
the one hand and new statistical and machine learning methods and
algorithms on the other. Many of these methods are computationally
intensive, and their performance can be significantly enhanced through
optimization. Ensemble techniques, which combine multiple models to
improve performance, further emphasize the need for computational power,
as they involve estimating diverse models and combining them in
meaningful ways.

Over the years, the
\href{https://CRAN.R-project.org/package=Rcpp}{\texorpdfstring%
{{\normalfont\fontseries{b}\selectfont Rcpp}}%
{Rcpp}} package has become a popular tool for R users to write and
integrate C++ code into R \citep{Rcpp}. This is particularly valuable
for computationally demanding tasks since C++ code is generally faster
than R code. However, writing and debugging C++ code has its challenges.
It can be difficult to determine whether the effort spent optimizing C++
code results in sufficient performance gains. The
\href{https://CRAN.R-project.org/package=rcpptimer}{\texorpdfstring%
{{\normalfont\fontseries{b}\selectfont rcpptimer}}%
{rcpptimer}} package addresses this by providing a straightforward tool
for benchmarking parallelized
\href{https://CRAN.R-project.org/package=Rcpp}{\texorpdfstring%
{{\normalfont\fontseries{b}\selectfont Rcpp}}%
{Rcpp}} code.

In this paper, we introduce rcpptimer and position it within the
existing range of profiling and benchmarking tools in R and C++. While
focusing on R, we briefly explore potential contributions to other
programming languages. The remainder of this introduction presents an
overview of current profiling and benchmarking tools in R and C++ and
shows the gap that
\href{https://CRAN.R-project.org/package=rcpptimer}{\texorpdfstring%
{{\normalfont\fontseries{b}\selectfont rcpptimer}}%
{rcpptimer}} closes.

Before discussing specific tools, it makes sense to distinguish between
profiling and benchmarking. Profiling is the process of measuring a
program's performance and identifying bottlenecks, while benchmarking
compares the performance of different implementations of the same task
\citep{wickham2019advanced}. These tasks often overlap in practice, and
some tools may be suitable for both.

In R, several functions and packages are available for profiling and
benchmarking. The simplest option is R's built-in \texttt{system.time()}
function, which measures the execution time of a single expression. The
bench package provides a high-precision equivalent,
\texttt{system\_time()} \citep{bench}. Beyond these, several more
advanced tools exist.

For example, the
\href{https://CRAN.R-project.org/package=rbenchmark}{\texorpdfstring%
{{\normalfont\fontseries{b}\selectfont rbenchmark}}%
{rbenchmark}} package wraps \texttt{system.time()} \citep{rbenchmark}.
It allows measuring multiple expressions. The expressions are repeatedly
evaluated, and the results are written to a data frame. However, the
CRAN version of rbenchmark is now 12 years old, and the package does not
seem to be under active development anymore\footnote{https://github.com/eddelbuettel/rbenchmark/commits/master/}.
Furthermore,
\href{https://CRAN.R-project.org/package=microbenchmark}{\texorpdfstring%
{{\normalfont\fontseries{b}\selectfont microbenchmark}}%
{microbenchmark}} and
\href{https://CRAN.R-project.org/package=bench}{\texorpdfstring%
{{\normalfont\fontseries{b}\selectfont bench}}%
{bench}} have practically the same interface but offer broader features.

\href{https://CRAN.R-project.org/package=microbenchmark}{\texorpdfstring%
{{\normalfont\fontseries{b}\selectfont microbenchmark}}%
{microbenchmark}} allows high precision timing of multiple expressions,
evaluating them 100 times by default \citep{microbenchmark}. The results
get written to a DataFrame. A print method is available to print the
results in a concise format. This method also provides a summary of the
results, including the mean and the median of the timings.
Microbenchmark is particularly useful for comparing the performance of
different implementations of the same small task, and it offers plot and
boxplot methods that are available for visualizing the results.

The \href{https://CRAN.R-project.org/package=bench}{\texorpdfstring%
{{\normalfont\fontseries{b}\selectfont bench}}%
{bench}} package has a very similar interface to
\href{https://CRAN.R-project.org/package=microbenchmark}{\texorpdfstring%
{{\normalfont\fontseries{b}\selectfont microbenchmark}}%
{microbenchmark}} \citep{bench}. It is also tailored for comparing
different implementations of the same small task; So much that
\href{https://CRAN.R-project.org/package=bench}{\texorpdfstring%
{{\normalfont\fontseries{b}\selectfont bench}}%
{bench}}s main function \texttt{mark()} throws an error if the results
of the expressions differ. The package evaluates the expressions until a
specified time limit or iteration cap is reached. Besides time, bench
also tracks memory usage and garbage collections and can run the
expressions on a large grid of values. Before summarizing the results,
iterations with garbace collections are filetered out. Like
\href{https://CRAN.R-project.org/package=microbenchmark}{\texorpdfstring%
{{\normalfont\fontseries{b}\selectfont microbenchmark}}%
{microbenchmark}},
\href{https://CRAN.R-project.org/package=bench}{\texorpdfstring%
{{\normalfont\fontseries{b}\selectfont bench}}%
{bench}} provides methods for printing and visualizing the results.

While \href{https://CRAN.R-project.org/package=bench}{\texorpdfstring%
{{\normalfont\fontseries{b}\selectfont bench}}%
{bench}} and
\href{https://CRAN.R-project.org/package=microbenchmark}{\texorpdfstring%
{{\normalfont\fontseries{b}\selectfont microbenchmark}}%
{microbenchmark}} are particularly designed for benchmarking, there are
also packages for profiling in R. One of these is
\href{https://CRAN.R-project.org/package=profvis}{\texorpdfstring%
{{\normalfont\fontseries{b}\selectfont profvis}}%
{profvis}} \citep{profvis}.
\href{https://CRAN.R-project.org/package=profvis}{\texorpdfstring%
{{\normalfont\fontseries{b}\selectfont profvis}}%
{profvis}} clearly focuses on profiling your R code. It provides an
interactive HTML-based visualization of profiling results, displayed as
a flame graph. Besides the timings,
\href{https://CRAN.R-project.org/package=profvis}{\texorpdfstring%
{{\normalfont\fontseries{b}\selectfont profvis}}%
{profvis}} also reports the memory usage of the code.
\href{https://CRAN.R-project.org/package=profvis}{\texorpdfstring%
{{\normalfont\fontseries{b}\selectfont profvis}}%
{profvis}} is particularly useful for identifying bottlenecks in your
code. A limitation of
\href{https://CRAN.R-project.org/package=profvis}{\texorpdfstring%
{{\normalfont\fontseries{b}\selectfont profvis}}%
{profvis}} is that it does not give insights into code that is
implemented in languages other than R (e.g.~C, C++, or Fortran).

Another distinct package that is worth mentioning is
\href{https://CRAN.R-project.org/package=tictoc}{\texorpdfstring%
{{\normalfont\fontseries{b}\selectfont tictoc}}%
{tictoc}} \citep{tictoc}. In contrast to the above, tictoc is designed
to integrate into your existing code. It is a simple timer that can be
used to measure the execution time of specific parts of your code.
Timers are started by calling \texttt{tic()} and stopped by calling
\texttt{toc()}. This simple interface makes
\href{https://CRAN.R-project.org/package=tictoc}{\texorpdfstring%
{{\normalfont\fontseries{b}\selectfont tictoc}}%
{tictoc}} suitable for both benchmarking and profiling. This simplicity
brings a big advantage: users do not need to extract parts out of
functions and possibly prepare input variables so that these parts can
be run independently. Instead, \texttt{tic()} and \texttt{toc()} calls
can be directly placed into existing code. This simplicity saves time
and creates incentives to actually use that package.
\href{https://CRAN.R-project.org/package=tictoc}{\texorpdfstring%
{{\normalfont\fontseries{b}\selectfont tictoc}}%
{tictoc}} is able to handle nested timers. However, overlapping timers
are not possible at the moment.

However, a growing number of R packages use
\href{https://CRAN.R-project.org/package=Rcpp}{\texorpdfstring%
{{\normalfont\fontseries{b}\selectfont Rcpp}}%
{Rcpp}} for calling C++ code from R. In the past, there were two main
options for measuring execution time of your C++ code.

The first option is using the toolbox available in R. This has multiple
disadvantages. The most obvious one is that one does not get insights
into the the execution time of specific parts of the C++ code. To solve
this, the parts of interest need to be isolated, which is a
time-consuming task. Furthermore, Measuring C++ execution Time in R
allways means measuring the execution time of a whole function,
including overhead that is created by passing input data to that
function.

The second option is using the toolbox available in C++. These options
include the chrono library, which is part of the C++ standard
library\footnote{https://en.cppreference.com/w/cpp/header/chrono},
Google Benchmark which is a library for microbenchmarking \footnote{https://github.com/google/benchmark?tab=readme-ov-file},
Celero \footnote{https://github.com/DigitalInBlue/Celero} which support
parameterized benchmarks, catch2 is a testing library that also provides
microbenchmarking capabilities \footnote{https://github.com/catchorg/Catch2},
and the boost library which also has a timer functionality \footnote{https://www.boost.org/doc/libs/1\_85\_0/libs/timer/doc/index.html}.
Those options are certainly the best in terms of C++ benchmarking and
profiling capabilities. However, using these tools means learning how to
use the them, adding external libraries to the code which may need to be
removed later, and writing code for passing the results to R
(i.e.~aggregating the results and altering the original return
statement). This is a time-consuming task, especially for
\href{https://CRAN.R-project.org/package=Rcpp}{\texorpdfstring%
{{\normalfont\fontseries{b}\selectfont Rcpp}}%
{Rcpp}} users who are not proficient in writing C++ code.

The rcpptimer package bridges the gap between these two approaches.
Inspired by tictoc, it brings a similar functionality to C++ code
integrated via
\href{https://CRAN.R-project.org/package=Rcpp}{\texorpdfstring%
{{\normalfont\fontseries{b}\selectfont Rcpp}}%
{Rcpp}}, while adding support for OpenMP parallelism
\citep{dagum1998openmp}. Built on RcppClock \citep{RcppClock},
\href{https://CRAN.R-project.org/package=rcpptimer}{\texorpdfstring%
{{\normalfont\fontseries{b}\selectfont rcpptimer}}%
{rcpptimer}} improves upon its predecessor by supporting OpenMP
parralelism and offering enhanced functionality, which will be discussed
in detail below.

The package fills a gap in the current range of benchmarking packages in
R and C++ as it is very simple, yet it enables users to benchmark
specific parts of their C++ code. The results are returned to R
automatically. Users do not need to write a single line of code for
this. The package supports OpenMP parallelism, nested and overlapping
timers, provides summary statistics, and safeguards users with
expressive warnings if necessary. Multiple vignettes thoroughly
demonstrate the usage of
\href{https://CRAN.R-project.org/package=rcpptimer}{\texorpdfstring%
{{\normalfont\fontseries{b}\selectfont rcpptimer}}%
{rcpptimer}}. The only dependence of
\href{https://CRAN.R-project.org/package=rcpptimer}{\texorpdfstring%
{{\normalfont\fontseries{b}\selectfont rcpptimer}}%
{rcpptimer}} is
\href{https://CRAN.R-project.org/package=Rcpp}{\texorpdfstring%
{{\normalfont\fontseries{b}\selectfont Rcpp}}%
{Rcpp}}. Furthermore, we regularly test the package with
\href{https://CRAN.R-project.org/package=testthat}{\texorpdfstring%
{{\normalfont\fontseries{b}\selectfont testthat}}%
{testthat}} and
\href{https://CRAN.R-project.org/package=rcmdcheck}{\texorpdfstring%
{{\normalfont\fontseries{b}\selectfont rcmdcheck}}%
{rcmdcheck}} on multiple platforms using GitHub actions.

To summarize, our main contrubtions with
\href{https://CRAN.R-project.org/package=rcpptimer}{\texorpdfstring%
{{\normalfont\fontseries{b}\selectfont rcpptimer}}%
{rcpptimer}} are:

\begin{itemize}
\tightlist
\item
  Bridging the gap between R and C++ tools for benchmarking and
  profiling
\item
  Straightforward interface: can be learned in just a few minutes
\item
  Allows timing of individual C++ sections
\item
  Support for OpenMP parralelism
\item
  Support for overlapping and nested timers
\item
  Automatically passing the results to R
\item
  Providing summary statistics
\end{itemize}

\href{https://CRAN.R-project.org/package=rcpptimer}{\texorpdfstring%
{{\normalfont\fontseries{b}\selectfont rcpptimer}}%
{rcpptimer}} is available on CRAN and can be installed using
\texttt{install.packages("rcpptimer")}.

The remainder of this package is structured as follows.
\hyperref[usage]{Chapter 2} briefly reviews how this package can be
used. In \hyperref[implementation]{Chapter 3}, we discuss how the core
timing capabilities, the computation of summary statistics and the
handover to R were implemented. Selected aspects that go beyond the
implementation of core functionality, such as warnings, timing scopes,
and resetting the timer will be discussed in \hyperref[other]{Chapter
4}. \hyperref[conclusion]{Chapter 5} concludes.

\section{Using rcpptimer}\label{usage}

The package can be used even without proficient knowledge of
\href{https://CRAN.R-project.org/package=Rcpp}{\texorpdfstring%
{{\normalfont\fontseries{b}\selectfont Rcpp}}%
{Rcpp}}. The first attempts to write C++ code using
\href{https://CRAN.R-project.org/package=Rcpp}{\texorpdfstring%
{{\normalfont\fontseries{b}\selectfont Rcpp}}%
{Rcpp}} usually involve \texttt{Rcpp::SourceCpp} or even
\texttt{Rcpp::cppFunction} to compile single files or functions. We
chose sensible defaults tailored for these use cases. Additionally,
extensive vignettes explain how users can integrate
\href{https://CRAN.R-project.org/package=rcpptimer}{\texorpdfstring%
{{\normalfont\fontseries{b}\selectfont rcpptimer}}%
{rcpptimer}} into their own packages and how they want to deviate from
the defaults when doing so.

\subsection{\texorpdfstring{Basic Usage with
\texttt{Rcpp::cppFunction}}{Basic Usage with Rcpp::cppFunction}}\label{BasicUsage}

In this section, we briefly show the package interface. Consider the
following example:

\begin{Shaded}
\begin{Highlighting}[]
\NormalTok{Rcpp}\SpecialCharTok{::}\FunctionTok{cppFunction}\NormalTok{(}
  \AttributeTok{code =} \StringTok{"}
\StringTok{  void atan\_vec(NumericVector \&x) \{}
\StringTok{    Timer timer;}
\StringTok{  \#pragma omp parallel for}
\StringTok{    for (double \&e : x)}
\StringTok{    \{}
\StringTok{      timer.tic();}
\StringTok{      e = atan(e);}
\StringTok{      timer.toc();}
\StringTok{    \}}
\StringTok{    DataFrame results = timer.stop();}
\StringTok{  \}}
\StringTok{  "}\NormalTok{,}
  \AttributeTok{depends =} \StringTok{"rcpptimer"}\NormalTok{,}
  \AttributeTok{plugins =} \StringTok{"openmp"}
\NormalTok{)}
\NormalTok{x }\OtherTok{\textless{}{-}} \FunctionTok{rnorm}\NormalTok{(}\DecValTok{1000}\NormalTok{)}
\NormalTok{x\_atan }\OtherTok{\textless{}{-}} \FunctionTok{atan\_vec}\NormalTok{(x)}
\end{Highlighting}
\end{Shaded}

\texttt{atan\_vec()} takes a numeric vector and replaces all values of
that vector with the arctangent of the respective values. The function
is compiled using \texttt{Rcpp::sourceCpp()}. Measuring the execution
time of the \texttt{atan()} calls inside the loop is as simple as
creating a timer object using \texttt{Timer\ timer;}, starting it with
\texttt{timer.tic();} and stopping it using \texttt{timer.toc();}. The
timer class will automatically return the results to R upon destruction.
The default name for the results is \texttt{times}. We can print the
results to the console using \texttt{print(times)}.

\begin{Shaded}
\begin{Highlighting}[]
\FunctionTok{print}\NormalTok{(times)}
\end{Highlighting}
\end{Shaded}

\begin{verbatim}
##        Microseconds     SD   Min     Max Count
## tictoc       12.251 66.225 0.067 870.598  1000
\end{verbatim}

The above code will use \texttt{print.data.frame()} or
\texttt{print.rcpptimer} (if the S3 method is registered) to print the
results. Both methods are similar; however, \texttt{print.rcpptimer} may
scale the results if specific criteria are met, e.g.~show nanoseconds
instead of microseconds if all measurements are below one microsecond.
Aside from the printing, the results are always returned in microseconds
with three decimal places. That is, we get insights on a nanosecond
level. We report summary statistics when multiple timer instances are
used, like above. Those get rounded to the closest nanosecond. Note that
the above example uses OpenMP parallelism. The user does not need to
worry about this. The timers will automatically distinguish between
threads (see \hyperref[ompsupport]{Section 3.2}). Assessing whether the
parallelism in \texttt{atan\_vec()} reduces overall computation time is
left as an exercise for the reader.

We decided to add the \texttt{Times} class to the
\href{https://CRAN.R-project.org/package=Rcpp}{\texorpdfstring%
{{\normalfont\fontseries{b}\selectfont Rcpp}}%
{Rcpp}} namespace so that users can use the \texttt{Times} class in
their own code if they attach the
\href{https://CRAN.R-project.org/package=Rcpp}{\texorpdfstring%
{{\normalfont\fontseries{b}\selectfont Rcpp}}%
{Rcpp}} namespace. \texttt{Rcpp::cppFunction} attaches this namespace
automatically; in that case,
\href{https://CRAN.R-project.org/package=rcpptimer}{\texorpdfstring%
{{\normalfont\fontseries{b}\selectfont rcpptimer}}%
{rcpptimer}} only needs to be added to the \texttt{depends} argument.

Upon construction the user can provide up to two arguments. The first
argument \texttt{name} determines the variable name of the resulting
DataFrame that gets passed to R (this defaults to \texttt{times}, see
\hyperref[PassToR]{Section 3.4}). The second argument \texttt{verbose}
is a boolean that indicates whether warnings should be printed (see
\hyperref[warnings]{Section 4.1}).

We provide four constructors for the Timer class:

\begin{Shaded}
\begin{Highlighting}[]
\NormalTok{Timer}\OperatorTok{()} \OperatorTok{:}\NormalTok{ CppTimer}\OperatorTok{()} \OperatorTok{\{\}}
\NormalTok{Timer}\OperatorTok{(}\AttributeTok{const} \DataTypeTok{char} \OperatorTok{*}\NormalTok{name}\OperatorTok{)} \OperatorTok{:}\NormalTok{ CppTimer}\OperatorTok{()} \OperatorTok{\{} \KeywordTok{this}\OperatorTok{{-}\textgreater{}}\NormalTok{name }\OperatorTok{=}\NormalTok{ name}\OperatorTok{;} \OperatorTok{\}}
\NormalTok{Timer}\OperatorTok{(}\DataTypeTok{bool}\NormalTok{ verbose}\OperatorTok{)} \OperatorTok{:}\NormalTok{ CppTimer}\OperatorTok{(}\NormalTok{verbose}\OperatorTok{)} \OperatorTok{\{\}}
\NormalTok{Timer}\OperatorTok{(}\AttributeTok{const} \DataTypeTok{char} \OperatorTok{*}\NormalTok{name}\OperatorTok{,} \DataTypeTok{bool}\NormalTok{ verbose}\OperatorTok{)} \OperatorTok{:}\NormalTok{ CppTimer}\OperatorTok{(}\NormalTok{verbose}\OperatorTok{)} \OperatorTok{\{} \KeywordTok{this}\OperatorTok{{-}\textgreater{}}\NormalTok{name }\OperatorTok{=}\NormalTok{ name}\OperatorTok{;} \OperatorTok{\}}
\end{Highlighting}
\end{Shaded}

That is: users may provide no arguments, just one of them, or both.

If only a single timer is needed, users can omit the \texttt{tag} and
call \texttt{.tic()} and \texttt{.toc()} without arguments (as above).
The \texttt{tag} argument is used to distinguish between timers and
defaults to \texttt{tictoc}.

In most cases, users will want to measure multiple code blocks and
possibly also scopes. In that case, \texttt{tag} argument is necessary.
We will consider the gibbs sampler as a more elaborate example. This
example was originally discussed by Dirk Eddelbuettel in a blog
post\footnote{https://gallery.rcpp.org/articles/gibbs-sampler/} and also
got added to \citep{wickham2019advanced}. In this example, we measure
the execution time of the entire function, the initialization of the
matrix, and the outer and inner loop. The \texttt{Timer} class makes
this very easy.

\begin{Shaded}
\begin{Highlighting}[]
\NormalTok{Rcpp}\SpecialCharTok{::}\FunctionTok{cppFunction}\NormalTok{(}
  \AttributeTok{code =} \StringTok{\textquotesingle{}}
\StringTok{  NumericMatrix gibbs\_cpp(int N, int thin) \{}
\StringTok{    Timer timer("gibbs\_cpp\_times");}
\StringTok{    Timer::ScopedTimer scope(timer, "gibbs\_cpp");}
\StringTok{    timer.tic("make\_matrix");}
\StringTok{    NumericMatrix mat(N, 2);}
\StringTok{    timer.toc("make\_matrix");}
\StringTok{    double x = 0, y = 0;}
\StringTok{    for(int i = 0; i \textless{} N; i++) \{}
\StringTok{      timer.tic("outer\_loop");}
\StringTok{      for(int j = 0; j \textless{} thin; j++) \{}
\StringTok{        timer.tic("inner\_loop");}
\StringTok{        x = rgamma(1, 3, 1 / (y * y + 4))[0];}
\StringTok{        y = rnorm(1, 1 / (x + 1), 1 / sqrt(2 * (x + 1)))[0];}
\StringTok{        timer.toc("inner\_loop");}
\StringTok{      \}}
\StringTok{      mat(i, 0) = x;}
\StringTok{      mat(i, 1) = y;}
\StringTok{      timer.toc("outer\_loop");}
\StringTok{    \}}
\StringTok{    return(mat);}
\StringTok{  \}}
\StringTok{  \textquotesingle{}}\NormalTok{,}
  \AttributeTok{depends =} \StringTok{"rcpptimer"}
\NormalTok{)}
\NormalTok{x }\OtherTok{\textless{}{-}} \FunctionTok{gibbs\_cpp}\NormalTok{(}\DecValTok{100}\NormalTok{, }\DecValTok{100}\NormalTok{)}
\end{Highlighting}
\end{Shaded}

In the above example, we use the \texttt{ScopedTimer} class to measure
the execution time of the entire function. The \texttt{ScopedTimer}
class is a helper class that automatically calls \texttt{.tic()} upon
creation and \texttt{.toc()} upon destruction (see
\hyperref[ScopedTimer]{Section 4.2}). We used matching \texttt{tags} in
the \texttt{.tic()} and \texttt{.toc()} calls above, to distinguish the
timers measuring the execution time of the matrix initialization, the
inner loop, and the outer loop. This time, the results are returend to R
as a DatFrame named \texttt{gibbs\_cpp\_times}.

\begin{Shaded}
\begin{Highlighting}[]
\FunctionTok{print}\NormalTok{(gibbs\_cpp\_times)}
\end{Highlighting}
\end{Shaded}

\begin{verbatim}
##             Microseconds    SD      Min      Max Count
## gibbs_cpp       3956.690 0.000 3956.690 3956.690     1
## inner_loop         0.362 0.529    0.201   22.749 10000
## make_matrix        4.472 0.000    4.472    4.472     1
## outer_loop        39.470 8.895   28.129   68.978   100
\end{verbatim}

We see that the scoped timer was indeed terminated automatically. The
timers in the loops got called multiple times. For these timers, we get
summary statistics (mean, standard deviation, minimum and maximum). The
above example shows that timers placed in nested loops may be called
many times. Returning the entire duration vector to R would be a
memory-demanding task. Therefore, we calculate the summary statistics in
C++ (it is possible to obtain the raw data; see
{[}\hyperref[PassToR]{Section 3.4}{]}).

\subsection{Using RcppTimer in a
Package}\label{using-rcpptimer-in-a-package}

Using rcpptimer in an R package is straightforward. First,
\href{https://CRAN.R-project.org/package=rcpptimer}{\texorpdfstring%
{{\normalfont\fontseries{b}\selectfont rcpptimer}}%
{rcpptimer}} needs to be added to the ``LinkingTo'' field in the
DESCRIPTION:

\begin{Shaded}
\begin{Highlighting}[]
\NormalTok{LinkingTo}\SpecialCharTok{:} \FunctionTok{rcpptimer}\NormalTok{ (}\SpecialCharTok{\textgreater{}=} \DecValTok{1}\NormalTok{.}\FloatTok{2.0}\NormalTok{)}
\end{Highlighting}
\end{Shaded}

The header file can now be included in the C++ scripts using
\texttt{\#include\ \textless{}rcpptimer.h\textgreater{}}. The Timer
class can then be used as in the examples above.

As a package author, it is often preferable to manage the timing results
within your code, rather than automatically assigning them to the user's
global environment. Package users probably do not expect your function
to write the \texttt{times} object into their global environment, and
may already have an object called \texttt{times} that would get
overwritten. Handling the results yourself is a straightforward task. We
modify the \texttt{atan\_vec()} function from above to demonstrate this.
Two steps are required: first, set \texttt{autoreturn} autoreturn to
\texttt{false}. Second, call the \texttt{.stop()} method of the
\texttt{Timer} class to obtain the results. \texttt{.stop()} returns the
exact DataFrame that would be returned to R if \texttt{autoreturn} was
set to true. The following code demonstrates this:

\begin{Shaded}
\begin{Highlighting}[]
\NormalTok{Rcpp}\SpecialCharTok{::}\FunctionTok{cppFunction}\NormalTok{(}
  \AttributeTok{code =} \StringTok{"}
\StringTok{  DataFrame atan\_vec(NumericVector \&x) \{}
\StringTok{    Timer timer;}
\StringTok{    timer.autoreturn = false;}
\StringTok{  \#pragma omp parallel for}
\StringTok{    for (double \&e : x)}
\StringTok{    \{}
\StringTok{      timer.tic();}
\StringTok{      e = atan(e);}
\StringTok{      timer.toc();}
\StringTok{    \}}
\StringTok{    DataFrame results = timer.stop();}
\StringTok{    return(results);}
\StringTok{  \}}
\StringTok{  "}\NormalTok{,}
  \AttributeTok{depends =} \StringTok{"rcpptimer"}
\NormalTok{)}
\NormalTok{x }\OtherTok{\textless{}{-}} \FunctionTok{rnorm}\NormalTok{(}\DecValTok{1000}\NormalTok{)}
\NormalTok{atac\_vec\_times }\OtherTok{\textless{}{-}} \FunctionTok{atan\_vec}\NormalTok{(x)}
\FunctionTok{print}\NormalTok{(atac\_vec\_times)}
\end{Highlighting}
\end{Shaded}

\begin{verbatim}
##        Nanoseconds SD Min  Max Count
## tictoc          75 55  58 1699  1000
\end{verbatim}

\texttt{.stop()} can be called multiple times. If new durations get
added in the meantime, it will update the results, otherwise, it will
just return the same DataFrame as before (See
\hyperref[SummaryStatistics]{Section 3.3}). In practice, you may want to
add the results DataFrame to a List and return it along with other
results. This is a common practice in R packages.

The autoreturn feature can theoretically also be used in conjunction
with manually calling \texttt{.stop()} this is discussed in the
\href{https://rcpptimer.berrisch.biz/articles/autoreturn.html}{Automatic
and Manual Return of the Timings} vignette.

Additionally, rcpptimer provides two functions, \texttt{fibonacci()} and
\texttt{fibonacci\_omp()}, as practical demonstrations\footnote{https://github.com/BerriJ/rcpptimer/blob/main/src/fibonacci.cpp}.
The latter showcases the use of rcpptimer in parallelized code. For the
sake of brevity, we will not delve into further details here and will
instead focus on the package's implementation specifics.

\section{Implementation details}\label{implementation}

This subsection provides an overview of the general structure of the
library. Originally, the package contained a single class
\texttt{Rcpp::Timer} which implemented the timing functionality and the
interface to R. In Februrary 2024 ((see commit
\href{https://github.com/BerriJ/rcpptimer/commit/babb8bcf2ddd2c6368d9b5f55ef7b40749b81350}{babb8bc})),
we split this class into an R-Specific part, which remained in
\href{https://CRAN.R-project.org/package=rcpptimer}{\texorpdfstring%
{{\normalfont\fontseries{b}\selectfont rcpptimer}}%
{rcpptimer}} and a generic class called \texttt{CppTimer} which we moved
to a standalone project called \texttt{cpptimer}\footnote{https://github.com/BerriJ/cpptimer}.
For brewity, we will drop the \texttt{Rcpp::} namespace qualifier of
\texttt{Rcpp::Timer} and refer to that class as the \texttt{Timer}
class.

Below, we show how \texttt{CppTimer} is structured. This overview is
simplified to highlight the most important aspects. In particular, we
removed most function bodies and include statements. For those
interested in a more detailed exploration, the complete source is
available on GitHub\footnote{https://github.com/BerriJ/rcpptimer/tree/main/inst/include}.

\begin{Shaded}
\begin{Highlighting}[]
\CommentTok{// cpptimer.h}

\KeywordTok{using} \KeywordTok{namespace}\NormalTok{ std}\OperatorTok{;}
\KeywordTok{using} \KeywordTok{namespace}\NormalTok{ chrono}\OperatorTok{;}

\KeywordTok{using}\NormalTok{ keypair }\OperatorTok{=}\NormalTok{ pair}\OperatorTok{\textless{}}\NormalTok{string}\OperatorTok{,} \DataTypeTok{unsigned} \DataTypeTok{int}\OperatorTok{\textgreater{};}
\KeywordTok{using}\NormalTok{ statistics }\OperatorTok{=}\NormalTok{ tuple}\OperatorTok{\textless{}}\DataTypeTok{double}\OperatorTok{,} \DataTypeTok{double}\OperatorTok{,} \DataTypeTok{double}\OperatorTok{,} \DataTypeTok{double}\OperatorTok{,} \DataTypeTok{unsigned} \DataTypeTok{long} \DataTypeTok{int}\OperatorTok{\textgreater{};}

\KeywordTok{class}\NormalTok{ CppTimer}
\OperatorTok{\{}
\KeywordTok{protected}\OperatorTok{:}
\NormalTok{  map}\OperatorTok{\textless{}}\NormalTok{keypair}\OperatorTok{,}\NormalTok{ high\_resolution\_clock}\OperatorTok{::}\NormalTok{time\_point}\OperatorTok{\textgreater{}}\NormalTok{ tics}\OperatorTok{;} \CommentTok{// Map of start times}
\NormalTok{  set}\OperatorTok{\textless{}}\NormalTok{string}\OperatorTok{\textgreater{}}\NormalTok{ missing\_tics}\OperatorTok{,}\NormalTok{ needless\_tocs}\OperatorTok{;}              \CommentTok{// Set of missing tics}
  \CommentTok{// Summary statistics to be returned: Tag, Mean, SD, Min, Max, Count}
\NormalTok{  map}\OperatorTok{\textless{}}\NormalTok{string}\OperatorTok{,}\NormalTok{ statistics}\OperatorTok{\textgreater{}}\NormalTok{ data}\OperatorTok{;}

\KeywordTok{public}\OperatorTok{:}
\NormalTok{  vector tags}\OperatorTok{;}                                \CommentTok{// Vector of identifiers}
\NormalTok{  vector durations}\OperatorTok{;}                           \CommentTok{// Vector of durations}
  \DataTypeTok{bool}\NormalTok{ verbose }\OperatorTok{=} \KeywordTok{true}\OperatorTok{;}                        \CommentTok{// Print warnings}
 
  \KeywordTok{template} \OperatorTok{\textless{}}\KeywordTok{typename}\NormalTok{ T}\OperatorTok{\textgreater{}}                       \CommentTok{// Constructors}
\NormalTok{  CppTimer}\OperatorTok{(}\NormalTok{T }\OperatorTok{\&\&)} \OperatorTok{=} \KeywordTok{delete}\OperatorTok{;}      
\NormalTok{  CppTimer}\OperatorTok{()}                  
\NormalTok{  CppTimer}\OperatorTok{(}\DataTypeTok{bool}\NormalTok{ verbose}\OperatorTok{)}      

  \DataTypeTok{void}\NormalTok{ toc}\OperatorTok{(}\NormalTok{string }\OperatorTok{\&\&}\NormalTok{tag }\OperatorTok{=} \StringTok{"tictoc"}\OperatorTok{)\{...\}}      \CommentTok{// Start timer}
  \DataTypeTok{void}\NormalTok{ tic}\OperatorTok{(}\NormalTok{string }\OperatorTok{\&\&}\NormalTok{tag }\OperatorTok{=} \StringTok{"tictoc"}\OperatorTok{)\{...\}}      \CommentTok{// Stop timer    }
  \KeywordTok{class}\NormalTok{ ScopedTimer }\OperatorTok{\{}                         \CommentTok{// Time Scopes}
  \KeywordTok{private}\OperatorTok{:}
\NormalTok{    CppTimer }\OperatorTok{\&}\NormalTok{timer}\OperatorTok{;}
\NormalTok{    string tag}\OperatorTok{;}
  \KeywordTok{public}\OperatorTok{:}
\NormalTok{    ScopedTimer}\OperatorTok{(}\NormalTok{CppTimer }\OperatorTok{\&}\NormalTok{timer}\OperatorTok{,}\NormalTok{ tag }\OperatorTok{=} \StringTok{"scoped"}\OperatorTok{)\{...\}}
    \OperatorTok{\textasciitilde{}}\NormalTok{ScopedTimer}\OperatorTok{()\{...\}}
  \OperatorTok{\}}   
\NormalTok{  map aggregate}\OperatorTok{()\{...\}}                        \CommentTok{// Calculate summary statistics}
  \DataTypeTok{void}\NormalTok{ reset}\OperatorTok{()}                                \CommentTok{// Reset the timer}
\OperatorTok{\}}
\end{Highlighting}
\end{Shaded}

The \texttt{CppTimer} class provides the core timing functionality. It
includes two constructors and holds member variables for storing timing
data, such as start times, durations, tags, and summary statistics.
Additionally, it contains members to keep track of unexpected behaviour.
The class offers methods to start and stop timers: \texttt{.tic()},
\texttt{.toc()}, and the \texttt{ScopedTimer} subclass for timing
scopes. It also includes management functions: \texttt{aggregate()}
calculates summary statistics and \texttt{reset()} resets the
\texttt{CppTimer} instance.

The \texttt{Timer} class, on the other hand, extends this functionality
to interact with R. Below is a simplified version of its implementation.

\begin{Shaded}
\begin{Highlighting}[]
\CommentTok{// rcpptimer.h}
\PreprocessorTok{\#include }\ImportTok{\textless{}Rcpp.h\textgreater{}}
\PreprocessorTok{\#include }\ImportTok{\textless{}cpptimer/cpptimer.h\textgreater{}}

\KeywordTok{namespace}\NormalTok{ Rcpp}
\OperatorTok{\{}
  \KeywordTok{class}\NormalTok{ Timer }\OperatorTok{:} \KeywordTok{public}\NormalTok{ CppTimer}
  \OperatorTok{\{}
  \KeywordTok{public}\OperatorTok{:}
    \BuiltInTok{std::}\NormalTok{string name }\OperatorTok{=} \StringTok{"times"}\OperatorTok{;}     \CommentTok{// Name DataFrame}
    \DataTypeTok{bool}\NormalTok{ autoreturn }\OperatorTok{=} \KeywordTok{true}\OperatorTok{;}

\NormalTok{    Timer}\OperatorTok{()}                         \CommentTok{// Constructors}
\NormalTok{    Timer}\OperatorTok{(}\AttributeTok{const} \DataTypeTok{char} \OperatorTok{*}\NormalTok{name}\OperatorTok{)}
\NormalTok{    Timer}\OperatorTok{(}\DataTypeTok{bool}\NormalTok{ verbose}\OperatorTok{)}
\NormalTok{    Timer}\OperatorTok{(}\AttributeTok{const} \DataTypeTok{char} \OperatorTok{*}\NormalTok{name}\OperatorTok{,} \DataTypeTok{bool}\NormalTok{ verbose}\OperatorTok{)}

    \DataTypeTok{void}\NormalTok{ print\_warnings}\OperatorTok{()\{} \OperatorTok{...} \OperatorTok{\}}    \CommentTok{// Print warnings}

\NormalTok{    DataFrame stop}\OperatorTok{()\{}               \CommentTok{// Pass data to R}
\NormalTok{      aggregate}\OperatorTok{();}
      \OperatorTok{...}
    \OperatorTok{\}}

    \OperatorTok{\textasciitilde{}}\NormalTok{Timer}\OperatorTok{()}
    \OperatorTok{\{}
      \ControlFlowTok{if} \OperatorTok{(}\NormalTok{autoreturn}\OperatorTok{)}
\NormalTok{        stop}\OperatorTok{();}
      \ControlFlowTok{if} \OperatorTok{(}\NormalTok{verbose}\OperatorTok{)}
\NormalTok{        print\_warnings}\OperatorTok{();}
    \OperatorTok{\}}
  \OperatorTok{\}}
\OperatorTok{\}}
\end{Highlighting}
\end{Shaded}

The \texttt{Timer} class extends the \texttt{CppTimer} class, inheriting
all its members and methods while adding R-specific functionality.
Specifically, it introduces constructors that allow users set the
\texttt{name} and set \texttt{verbose} to false when desired. The class
also features a \texttt{print\_warnings()} method to display warnings
when necessary. The \texttt{stop()} method calculates summary statistics
by invoking \texttt{aggregate()} and returns the results to R as a
DataFrame. In its destructor, the Timer class may automatically call
\texttt{print\_warnings()} and \texttt{stop()} if \texttt{verbose} and
\texttt{autoreturn} are set to \texttt{true}.

The following subsections provide a more detailed discussion of
implementing the \texttt{CppTimer} and \texttt{Timer} classes.

\subsection{Starting and Stopping
Timers}\label{starting-and-stopping-timers}

Arguably, the two most importnat methods of the \texttt{Timer} class are
\texttt{.tic()} and \texttt{.toc()}. These methods start and stop the
timers, respectively. The implementation of these two methods follows
two goals: first, they should be as fast as possible so that the
overhead of the timers is negligible. Second, they should be as simple
to use as possible. We took two measures to achieve user friendlyness.
First, we renamed \texttt{.tick()} and \texttt{.tock()} to
\texttt{.tic()} and \texttt{.toc()} to align the interface to the tictoc
R package\footnote{https://github.com/BerriJ/rcpptimer/commit/e65e3fea07f83e407d305455df44f7335182fdb3}.
Second, we added a default argument to both methods so they can be
called without arguments. This is convenient when only a single timer is
needed.

Ultimately, we are interested in measuring the time differences between
matching \texttt{.tic()} and \texttt{.toc()} calls. There are several
ways to calculate these durations. In the initial implementation of
\href{https://CRAN.R-project.org/package=RcppClock}{\texorpdfstring%
{{\normalfont\fontseries{b}\selectfont RcppClock}}%
{RcppClock}}, \texttt{.tick()} and \texttt{.tock()} pushed data into
vectors, and the durations were calculated afterward. However, matching
the duration vectors was computationally intensive as overlapping timers
were possible and therefore, the duration entries needed to be matched
using tag vectors. This task was time-consuming as it involed costly
string comparisons and \texttt{.erase()} operations. To mitigate this
problem,
\href{https://CRAN.R-project.org/package=RcppClock}{\texorpdfstring%
{{\normalfont\fontseries{b}\selectfont RcppClock}}%
{RcppClock}} moved the matching algorithm into the \texttt{.tock()}
method in 2021\footnote{https://github.com/BerriJ/rcpptimer/commit/aa19372520dcc28125e3b6af01a0ada2aa4fa1d0}.

We take a different approach that is fast, memory efficient, and
designed to support OpenMP parralelism. To illustrate this idea, first
consider a simplified example without OpenMP support. Below is the
\texttt{.tic()} implementation:

\begin{Shaded}
\begin{Highlighting}[]
\KeywordTok{protected}\OperatorTok{:}
  \BuiltInTok{std::}\NormalTok{map}\OperatorTok{\textless{}}\BuiltInTok{std::}\NormalTok{string}\OperatorTok{,} \BuiltInTok{std::}\NormalTok{chrono::high\_resolution\_clock::time\_point}\OperatorTok{\textgreater{}}\NormalTok{ tics}\OperatorTok{;}

\DataTypeTok{void}\NormalTok{ tic}\OperatorTok{(}\BuiltInTok{std::}\NormalTok{string }\OperatorTok{\&\&}\NormalTok{tag }\OperatorTok{=} \StringTok{"tictoc"}\OperatorTok{)}
\OperatorTok{\{}
\NormalTok{  tics}\OperatorTok{[}\NormalTok{tag}\OperatorTok{]} \OperatorTok{=} \BuiltInTok{std::}\NormalTok{chrono::high\_resolution\_clock::now}\OperatorTok{();}
\OperatorTok{\}}
\end{Highlighting}
\end{Shaded}

To store the exact time point \texttt{.tic()} is called we use
\texttt{std::chrono::high\_resolution\_clock} which is an alias to the
clock with the highest precision available\footnote{https://en.cppreference.com/w/cpp/chrono/high\_resolution\_clock}.

This implementation uses \texttt{tics} (a \texttt{std::map} and
protected member of \texttt{CppTimer}) to store the time points. The
map's key is the tag of the timer (a \texttt{std::string}, usually
supplied by the user). That is, the \texttt{tag} argument is used to
distinguish between timers. Implementing an appropriate \texttt{.toc()}
method is straightfowrd. Consider a simplified version below:

\begin{Shaded}
\begin{Highlighting}[]
\DataTypeTok{void}\NormalTok{ toc}\OperatorTok{(}\BuiltInTok{std::}\NormalTok{string }\OperatorTok{\&\&}\NormalTok{tag}\OperatorTok{)}
\OperatorTok{\{}
  \BuiltInTok{std::}\NormalTok{chrono::nanoseconds duration }\OperatorTok{=} 
      \BuiltInTok{std::}\NormalTok{chrono::high\_resolution\_clock::now}\OperatorTok{()} \OperatorTok{{-}}\NormalTok{ tics}\OperatorTok{[}\NormalTok{tag}\OperatorTok{];}
\NormalTok{  durations}\OperatorTok{.}\NormalTok{push\_back}\OperatorTok{(}\NormalTok{duration}\OperatorTok{.}\NormalTok{count}\OperatorTok{());}
\NormalTok{  tags}\OperatorTok{.}\NormalTok{push\_back}\OperatorTok{(}\BuiltInTok{std::}\NormalTok{move}\OperatorTok{(}\NormalTok{tag}\OperatorTok{));}
\OperatorTok{\}}
\end{Highlighting}
\end{Shaded}

This version abstracts away OpenMP support and warning messages, which
we will address later. In this implementation, .toc() calculates the
duration by subtracting the time point stored in the tics map from the
current time. The correct time point is located using the corresponding
tag. Accessing the entry in the \texttt{tics} map is a cheap operation
with \(O(n\log{}n)\) complexity. The calculated duration is then stored
in the \texttt{durations} vector , and the associated tag is saved in
the tags vector. Using a map for tics, instead of storing all time
points in vectors, has an additional advantage: the map remains compact,
growing only to the number of unique tags used. When a timer is called
multiple times, the previous values in the map are overwritten, keeping
memory usage low.

The implementations of \texttt{.tic()} and \texttt{.toc()} above already
support overlapping timers, provided that each timer has a unique tag.
That is, an arbitrary number of other timers can be started and stopped
between starting and stopping a particular timer. The
\href{https://CRAN.R-project.org/package=tictoc}{\texorpdfstring%
{{\normalfont\fontseries{b}\selectfont tictoc}}%
{tictoc}} R package also supports this functionality. However,
\href{https://CRAN.R-project.org/package=rcpptimer}{\texorpdfstring%
{{\normalfont\fontseries{b}\selectfont rcpptimer}}%
{rcpptimer}} offers even more flexibility, it also supports overlapping
timers. This means that a second timer, started after the first, doesn't
need to stop before stopping the first one. Both features work because
\texttt{.tic()} stores time points along with their tags, enabling toc()
to match the stopping points correctly (see the
\href{https://rcpptimer.berrisch.biz/articles/rcpptimer.html}{introductory
vignette} for demonstrations). However, these implementations do not yet
support OpenMP parallelism. The next subsection explains how we
integrated OpenMP functionality into these methods.

\subsection{OMP Support}\label{ompsupport}

OpenMP is a popular and easy-to-use library for parallel programming in
C++. However, the implementations above are not suited for measuring
parallelized code. To explain this in more detail, consider the
\texttt{atan\_vec()} function from \hyperref[BasicUsage]{Section 2.1}
again:

\begin{Shaded}
\begin{Highlighting}[]
\DataTypeTok{void}\NormalTok{ atan\_vec}\OperatorTok{(}\NormalTok{NumericVector }\OperatorTok{\&}\NormalTok{x}\OperatorTok{)} \OperatorTok{\{}
\NormalTok{  Timer timer}\OperatorTok{;}
  \PreprocessorTok{\#pragma omp parallel for}
  \ControlFlowTok{for} \OperatorTok{(}\DataTypeTok{double} \OperatorTok{\&}\NormalTok{e }\OperatorTok{:}\NormalTok{ x}\OperatorTok{)}
  \OperatorTok{\{}
\NormalTok{    timer}\OperatorTok{.}\NormalTok{tic}\OperatorTok{(}\StringTok{"atan"}\OperatorTok{);}
\NormalTok{    e }\OperatorTok{=}\NormalTok{ atan}\OperatorTok{(}\NormalTok{e}\OperatorTok{);}
\NormalTok{    timer}\OperatorTok{.}\NormalTok{toc}\OperatorTok{(}\StringTok{"atan"}\OperatorTok{);}
  \OperatorTok{\}}
\OperatorTok{\}}
\end{Highlighting}
\end{Shaded}

The \texttt{\#pragma\ omp\ parallel\ for} tells the compiler to
parralelize the following. Multiple threads then execute the loop in
parallel, creating two problems for the implementations above.

The first is, that \texttt{.tic()} and \texttt{.toc()} do not yet
distinguish between threads. That is, the ``atan'' entries of the
\texttt{ticmap} will be overwritten by different threads. That is, we
cannot guarantee that the \texttt{.toc()} method will match the correct
\texttt{.tic()} call - it will simply match the most recent one,
regardless of the thread.

The second problem is that \texttt{.tic()} and \texttt{.toc()} will
write into \texttt{ticmap}, \texttt{durations} and \texttt{tags}
concurrently. This can create a data race where multiple threads attempt
to write to the same shared object. This leads to undefined behaviour
and can potentially cause a crash.

The following two subsections explain how we solve these issues.

\subsubsection{Distinguishing Between
Threads}\label{distinguishing-between-threads}

We solve the first problem by extending the key of the map. That is, we
use a \texttt{std::pair} containing the tag and the thread number,
instead of solely using the tag. This way, each thread gets its own
timers. The thread number is obtained via
\texttt{omp\_get\_thread\_num()}. This function is part of
\texttt{omp.h}. We validate if OpenMP available by checking whether
\texttt{\_OPENMP} is defined. If not, we define
\texttt{omp\_get\_thread\_num()} to allways return 0. This way, the code
still works without OpenMP. The following code snippet shows the
definition of \texttt{omp\_get\_thread\_num()}, the modified definition
of the \texttt{tics} map, and the \texttt{.tic()} method:

\begin{Shaded}
\begin{Highlighting}[]
\PreprocessorTok{\#ifndef \_OPENMP}
\KeywordTok{inline} \DataTypeTok{int}\NormalTok{ omp\_get\_thread\_num}\OperatorTok{()} \OperatorTok{\{} \ControlFlowTok{return} \DecValTok{0}\OperatorTok{;} \OperatorTok{\}}
\PreprocessorTok{\#endif}

\KeywordTok{using}\NormalTok{ keypair }\OperatorTok{=} \BuiltInTok{std::}\NormalTok{pair}\OperatorTok{\textless{}}\BuiltInTok{std::}\NormalTok{string}\OperatorTok{,} \DataTypeTok{unsigned} \DataTypeTok{int}\OperatorTok{\textgreater{};}
\KeywordTok{private}\OperatorTok{:}
  \BuiltInTok{std::}\NormalTok{map}\OperatorTok{\textless{}}\NormalTok{keypair}\OperatorTok{,}\NormalTok{ high\_resolution\_clock}\OperatorTok{::}\NormalTok{time\_point}\OperatorTok{\textgreater{}}\NormalTok{ tics}\OperatorTok{;}

\DataTypeTok{void}\NormalTok{ tic}\OperatorTok{(}\BuiltInTok{std::}\NormalTok{string }\OperatorTok{\&\&}\NormalTok{tag }\OperatorTok{=} \StringTok{"tictoc"}\OperatorTok{)}
\OperatorTok{\{}
\NormalTok{  keypair key}\OperatorTok{(}\BuiltInTok{std::}\NormalTok{move}\OperatorTok{(}\NormalTok{tag}\OperatorTok{),}\NormalTok{ omp\_get\_thread\_num}\OperatorTok{());}
\NormalTok{  tics}\OperatorTok{[}\NormalTok{key}\OperatorTok{]} \OperatorTok{=}\NormalTok{ hr\_clock}\OperatorTok{::}\NormalTok{now}\OperatorTok{();}
\OperatorTok{\}}
\end{Highlighting}
\end{Shaded}

The \texttt{.toc()} method was modified accordingly.

\subsubsection{Thread Safety}\label{thread-safety}

Parallel execution poses challenges for operations that modify shared
data. While reading elements from the \texttt{tics} map is safe since
\texttt{std::map} supports concurrent read access\footnote{https://en.cppreference.com/w/cpp/container},
writing to \texttt{tics}, as well as to \texttt{durations} and
\texttt{tags}, is problematic. We address this issue by employing OpenMP
critical regions. The following code snippet demonstrates the modified
\texttt{.toc()} method:

\begin{Shaded}
\begin{Highlighting}[]
\DataTypeTok{void}\NormalTok{ toc}\OperatorTok{(}\BuiltInTok{std::}\NormalTok{string }\OperatorTok{\&\&}\NormalTok{tag}\OperatorTok{)}
\OperatorTok{\{}
\NormalTok{  keypair key}\OperatorTok{(}\BuiltInTok{std::}\NormalTok{move}\OperatorTok{(}\NormalTok{tag}\OperatorTok{),}\NormalTok{ omp\_get\_thread\_num}\OperatorTok{());}
  \BuiltInTok{std::}\NormalTok{chrono::nanoseconds duration }\OperatorTok{=} 
      \BuiltInTok{std::}\NormalTok{chrono::high\_resolution\_clock::now}\OperatorTok{()} \OperatorTok{{-}}\NormalTok{ tics}\OperatorTok{[}\NormalTok{key}\OperatorTok{];}

  \PreprocessorTok{\#pragma omp critical\{  }
\NormalTok{  durations}\OperatorTok{.}\NormalTok{push\_back}\OperatorTok{(}\NormalTok{duration}\OperatorTok{.}\NormalTok{count}\OperatorTok{());}
\NormalTok{  tags}\OperatorTok{.}\NormalTok{push\_back}\OperatorTok{(}\BuiltInTok{std::}\NormalTok{move}\OperatorTok{(}\NormalTok{tag}\OperatorTok{));}
  \OperatorTok{\}}
\OperatorTok{\}}
\end{Highlighting}
\end{Shaded}

The \texttt{\#pragma\ omp\ critical} statement ensures that the
following block is executed by only one thread at a time, effectively
eliminating data races. This way, we avoid any issues w.r.t. writing to
the \texttt{durations} and \texttt{tags} vectors.We have also updated
the \texttt{.tic()} method accordingly. It is important to note that we
only pass the tag to the tags vector and do not include the thread
number. This is because we do not need to distinguish between threads in
the results; parallelized code will produce multiple entries in the
durations vector with the same tag.

In summary, we have addressed the two issues that arise when measuring
parallelized code. The implementation is now thread-safe, as the timers
distinguish between threads.

\subsection{Calculating Summary Statistics}\label{SummaryStatistics}

As discussed above, \texttt{.tic()} and \texttt{.toc()} work in together
to fill the \texttt{durations} and \texttt{tags} vectors. These vectors
can become quite large when timers are used within (parallelized) loops,
making it memory- and time-consuming to return them to R. Furthermore,
inspecting them may be cumbersome. To adress this, we calculate summary
statistics in C++ before returning the results to R. This is the done by
the \texttt{aggregate()} method. This method serves two main purposes:
first, it groups durations with the same tag together, and second, it
calculates statistics for each group.

There are various ways to tackle this task. Initially, we generated a
vector of unique \texttt{tags} using \texttt{std::sort} in combination
with \texttt{std::unique}. We then iterated over the unique
\texttt{tags}, and for each one, we looped through the entire
\texttt{tags} vector again to calculate summary statistics when
\texttt{tags} matched. This approach required costly string comparisons
and multiple iterations through the \texttt{tags} vector.

The current implementation is more efficient. It requires only a single
iteration over the tags vector, and by using a std::map to store the
results, we eliminate the need for string comparisons, allowing the
results to be grouped automatically. We calculate the summary statistics
online, updating them iteratively while traversing the tags vector. This
has several advantages: we do not need to create separate duration
vectors for each tag, and the statistics can be efficiently updated as
new durations are added, withouth evaluating the old durations again.

In total, we compute five properties for each timer: the mean, the
standard deviation, the minimum, the maximum, and the number of
measurements. The following code snippet presents the current
implementation of the \texttt{aggregate()} method:

\begin{Shaded}
\begin{Highlighting}[]
\NormalTok{map}\OperatorTok{\textless{}}\NormalTok{string}\OperatorTok{,}\NormalTok{ statistics}\OperatorTok{\textgreater{}}\NormalTok{ aggregate}\OperatorTok{()}
\OperatorTok{\{}
  \CommentTok{// Calculate summary statistics}
  \ControlFlowTok{for} \OperatorTok{(}\DataTypeTok{unsigned} \DataTypeTok{long} \DataTypeTok{int}\NormalTok{ i }\OperatorTok{=} \DecValTok{0}\OperatorTok{;}\NormalTok{ i }\OperatorTok{\textless{}}\NormalTok{ tags}\OperatorTok{.}\NormalTok{size}\OperatorTok{();}\NormalTok{ i}\OperatorTok{++)}
  \OperatorTok{\{}
    \CommentTok{// Welford\textquotesingle{}s online algorithm for mean and sst}
    \CommentTok{// sst = sum of squared total deviations}
    \DataTypeTok{double}\NormalTok{ mean }\OperatorTok{=} \DecValTok{0}\OperatorTok{,}\NormalTok{ sst }\OperatorTok{=} \DecValTok{0}\OperatorTok{,}\NormalTok{ min }\OperatorTok{=}\NormalTok{ numeric\_limits}\OperatorTok{\textless{}}\DataTypeTok{double}\OperatorTok{\textgreater{}::}\NormalTok{max}\OperatorTok{(),}\NormalTok{ max }\OperatorTok{=} \DecValTok{0}\OperatorTok{;}
    \DataTypeTok{unsigned} \DataTypeTok{long} \DataTypeTok{int}\NormalTok{ count }\OperatorTok{=} \DecValTok{0}\OperatorTok{;}
    \ControlFlowTok{if} \OperatorTok{(}\KeywordTok{auto}\NormalTok{ entry}\OperatorTok{\{}\NormalTok{data}\OperatorTok{.}\NormalTok{find}\OperatorTok{(}\NormalTok{tags}\OperatorTok{[}\NormalTok{i}\OperatorTok{])\};}\NormalTok{ entry }\OperatorTok{!=}\NormalTok{ end}\OperatorTok{(}\NormalTok{data}\OperatorTok{))}
    \OperatorTok{\{}
\NormalTok{      tie}\OperatorTok{(}\NormalTok{mean}\OperatorTok{,}\NormalTok{ sst}\OperatorTok{,}\NormalTok{ min}\OperatorTok{,}\NormalTok{ max}\OperatorTok{,}\NormalTok{ count}\OperatorTok{)} \OperatorTok{=}\NormalTok{ entry}\OperatorTok{{-}\textgreater{}}\NormalTok{second}\OperatorTok{;}
    \OperatorTok{\}}
\NormalTok{    count}\OperatorTok{++;}
    \DataTypeTok{double}\NormalTok{ duration }\OperatorTok{=}\NormalTok{ durations}\OperatorTok{[}\NormalTok{i}\OperatorTok{];}
    \DataTypeTok{double}\NormalTok{ delta }\OperatorTok{=}\NormalTok{ duration }\OperatorTok{{-}}\NormalTok{ mean}\OperatorTok{;}
\NormalTok{    mean }\OperatorTok{+=}\NormalTok{ delta }\OperatorTok{/}\NormalTok{ count}\OperatorTok{;}
\NormalTok{    sst }\OperatorTok{+=}\NormalTok{ delta }\OperatorTok{*} \OperatorTok{(}\NormalTok{duration }\OperatorTok{{-}}\NormalTok{ mean}\OperatorTok{);}
\NormalTok{    min }\OperatorTok{=} \BuiltInTok{std::}\NormalTok{min}\OperatorTok{(}\NormalTok{min}\OperatorTok{,}\NormalTok{ duration}\OperatorTok{);}
\NormalTok{    max }\OperatorTok{=} \BuiltInTok{std::}\NormalTok{max}\OperatorTok{(}\NormalTok{max}\OperatorTok{,}\NormalTok{ duration}\OperatorTok{);}
\NormalTok{    data}\OperatorTok{[}\NormalTok{tags}\OperatorTok{[}\NormalTok{i}\OperatorTok{]]} \OperatorTok{=} \OperatorTok{\{}\NormalTok{mean}\OperatorTok{,}\NormalTok{ sst}\OperatorTok{,}\NormalTok{ min}\OperatorTok{,}\NormalTok{ max}\OperatorTok{,}\NormalTok{ count}\OperatorTok{\};}
  \OperatorTok{\}}
\NormalTok{  tags}\OperatorTok{.}\NormalTok{clear}\OperatorTok{(),}\NormalTok{ durations}\OperatorTok{.}\NormalTok{clear}\OperatorTok{();}
  \ControlFlowTok{return} \OperatorTok{(}\NormalTok{data}\OperatorTok{);}
\OperatorTok{\}}
\end{Highlighting}
\end{Shaded}

The \texttt{data} map is a protected member of the \texttt{CppTimer}
class. It stores the summary statistics for each \texttt{tag}. The value
of the map is a \texttt{std::tuple} containing the five summary
statistics. While iterating through the \texttt{tags} vector, we check
if \texttt{data} aready contains an entry for the \texttt{tag}. If it
does, we initialize \texttt{mean}, \texttt{sst}, \texttt{min},
\texttt{max} and \texttt{count} using that entry. If it does not, we
initilize these variables with zero except \texttt{min} which is
initialized with
\texttt{numeric\_limits\textless{}double\textgreater{}::max()}, i.e.~a
very large constant. Then, we calculate the new summary statistics. We
use Welford's online algorithm to calculate the mean and the sum of
squared total deviations (SST) iteratively \citep{welford1962note}.
Finally, we update the entry of the \texttt{data} map.

This approach has several advantages. It iterates through the tags
vector only once, does not require costly string comparisons, and avoids
sorting the tags vector. The results are grouped automatically as we
utilizize the same map for storing the statistics and initializing the
algorithm. Finally, the summary statistics are calculated online,
i.e.~iteratively. This is efficient and allows updating the statistics
if new durations are added.

After iterating through \texttt{tags}, we clear the \texttt{tags} and
\texttt{durations} vectors. If a user calls \texttt{aggregate()} again,
summary statistics will be updated without having to evaluate the old
durations again. This is useful if a user wants to inspect the results
for a code segment executed again afterwards. In such cases, the user
can call \texttt{aggregate()}, review the results, and then, after the
code segment ran again, call \texttt{aggregate()} once more to obtain
the updated results. It is also possible to \texttt{reset()} the timer
instance (i.e.~remove all entries in \texttt{data}) so that the
statistics are calculated from scratch (see \hyperref[reset]{Section
4.2} ).

Finally, \texttt{aggregate()} returns data, which is useful if a user
wishes to call \texttt{aggregate()} manually. However, the default
behaviour of rcpptimer is to pass the results to R automatically. We
will discuss this next.

\subsection{Passing the results to R}\label{PassToR}

While\texttt{.tic()}, \texttt{.toc()}, and \texttt{aggregate()} are
members of the \texttt{CppTimer} class and are inherited by the
\texttt{Timer} class, the \texttt{.stop()} method contains R-specific
code and is, therefore, a class-specific method of the Timer class.

The \texttt{.stop()} method passes the results to \texttt{R}. This
involves several steps. The first is aggregating the data,
i.e.~calculating the summary statistics. This is done by calling the
\texttt{.aggregate()} method. The second step is preparing the data for
R. For our data, a DataFrame is the most suitable data structure. The
last step is returning the DataFrame to \texttt{R}. Finally, the method
returns the DataFrame to R. The following code snippet illustrates the
current implementation of the \texttt{.stop()} method:

\begin{Shaded}
\begin{Highlighting}[]
\NormalTok{DataFrame stop}\OperatorTok{()}
\OperatorTok{\{}
\NormalTok{  aggregate}\OperatorTok{(),}\NormalTok{ fesetround}\OperatorTok{(}\NormalTok{FE\_TONEAREST}\OperatorTok{);}

  \CommentTok{// Output Objects}
\NormalTok{  vector}\OperatorTok{\textless{}}\NormalTok{string}\OperatorTok{\textgreater{}}\NormalTok{ out\_tags}\OperatorTok{;}
\NormalTok{  vector}\OperatorTok{\textless{}}\DataTypeTok{unsigned} \DataTypeTok{long} \DataTypeTok{int}\OperatorTok{\textgreater{}}\NormalTok{ out\_counts}\OperatorTok{;}
\NormalTok{  vector}\OperatorTok{\textless{}}\DataTypeTok{double}\OperatorTok{\textgreater{}}\NormalTok{ out\_mean}\OperatorTok{,}\NormalTok{ out\_sd}\OperatorTok{,}\NormalTok{ out\_min}\OperatorTok{,}\NormalTok{ out\_max}\OperatorTok{;}

  \ControlFlowTok{for} \OperatorTok{(}\KeywordTok{auto} \AttributeTok{const} \OperatorTok{\&}\NormalTok{entry }\OperatorTok{:}\NormalTok{ data}\OperatorTok{)}
  \OperatorTok{\{}
\NormalTok{    out\_tags}\OperatorTok{.}\NormalTok{push\_back}\OperatorTok{(}\NormalTok{entry}\OperatorTok{.}\NormalTok{first}\OperatorTok{);}
    \KeywordTok{auto} \OperatorTok{[}\NormalTok{mean}\OperatorTok{,}\NormalTok{ sst}\OperatorTok{,}\NormalTok{ min}\OperatorTok{,}\NormalTok{ max}\OperatorTok{,}\NormalTok{ count}\OperatorTok{]} \OperatorTok{=}\NormalTok{ entry}\OperatorTok{.}\NormalTok{second}\OperatorTok{;}
    \CommentTok{// round to the nearest integer and to even in halfway cases and}
    \CommentTok{// convert to microseconds}
\NormalTok{    out\_mean}\OperatorTok{.}\NormalTok{push\_back}\OperatorTok{(}\NormalTok{nearbyint}\OperatorTok{(}\NormalTok{mean}\OperatorTok{)} \OperatorTok{*} \FloatTok{1e{-}3}\OperatorTok{);}
    \CommentTok{// Bessels\textquotesingle{} correction}
    \DataTypeTok{double}\NormalTok{ variance }\OperatorTok{=}\NormalTok{ sst }\OperatorTok{/} \BuiltInTok{std::}\NormalTok{max}\OperatorTok{(}\DataTypeTok{double}\OperatorTok{(}\NormalTok{count }\OperatorTok{{-}} \DecValTok{1}\OperatorTok{),} \FloatTok{1.0}\OperatorTok{);}
\NormalTok{    out\_sd}\OperatorTok{.}\NormalTok{push\_back}\OperatorTok{(}\NormalTok{nearbyint}\OperatorTok{(}\BuiltInTok{std::}\NormalTok{sqrt}\OperatorTok{(}\NormalTok{variance }\OperatorTok{*} \FloatTok{1e{-}6}\OperatorTok{)} \OperatorTok{*} \FloatTok{1e+3}\OperatorTok{)} \OperatorTok{*} \FloatTok{1e{-}3}\OperatorTok{);}
\NormalTok{    out\_min}\OperatorTok{.}\NormalTok{push\_back}\OperatorTok{(}\NormalTok{nearbyint}\OperatorTok{(}\NormalTok{min}\OperatorTok{)} \OperatorTok{*} \FloatTok{1e{-}3}\OperatorTok{);}
\NormalTok{    out\_max}\OperatorTok{.}\NormalTok{push\_back}\OperatorTok{(}\NormalTok{nearbyint}\OperatorTok{(}\NormalTok{max}\OperatorTok{)} \OperatorTok{*} \FloatTok{1e{-}3}\OperatorTok{);}
\NormalTok{    out\_counts}\OperatorTok{.}\NormalTok{push\_back}\OperatorTok{(}\NormalTok{count}\OperatorTok{);}
  \OperatorTok{\}}
\NormalTok{  DataFrame results }\OperatorTok{=}\NormalTok{ DataFrame}\OperatorTok{::}\NormalTok{create}\OperatorTok{(}
\NormalTok{      Named}\OperatorTok{(}\StringTok{"Microseconds"}\OperatorTok{)} \OperatorTok{=}\NormalTok{ out\_mean}\OperatorTok{,}
\NormalTok{      Named}\OperatorTok{(}\StringTok{"SD"}\OperatorTok{)} \OperatorTok{=}\NormalTok{ out\_sd}\OperatorTok{,}
\NormalTok{      Named}\OperatorTok{(}\StringTok{"Min"}\OperatorTok{)} \OperatorTok{=}\NormalTok{ out\_min}\OperatorTok{,}
\NormalTok{      Named}\OperatorTok{(}\StringTok{"Max"}\OperatorTok{)} \OperatorTok{=}\NormalTok{ out\_max}\OperatorTok{,}
\NormalTok{      Named}\OperatorTok{(}\StringTok{"Count"}\OperatorTok{)} \OperatorTok{=}\NormalTok{ out\_counts}\OperatorTok{);}
\NormalTok{  results}\OperatorTok{.}\NormalTok{attr}\OperatorTok{(}\StringTok{"class"}\OperatorTok{)} \OperatorTok{=}\NormalTok{ CharacterVector}\OperatorTok{(\{}\StringTok{"rcpptimer"}\OperatorTok{,} \StringTok{"data.frame"}\OperatorTok{\});}
\NormalTok{  results}\OperatorTok{.}\NormalTok{attr}\OperatorTok{(}\StringTok{"row.names"}\OperatorTok{)} \OperatorTok{=}\NormalTok{ out\_tags}\OperatorTok{;}
  \ControlFlowTok{if} \OperatorTok{(}\NormalTok{autoreturn}\OperatorTok{)}
  \OperatorTok{\{}
\NormalTok{    Environment env }\OperatorTok{=}\NormalTok{ Environment}\OperatorTok{::}\NormalTok{global\_env}\OperatorTok{();}
\NormalTok{    env}\OperatorTok{[}\NormalTok{name}\OperatorTok{]} \OperatorTok{=}\NormalTok{ results}\OperatorTok{;}
  \OperatorTok{\}}
  \ControlFlowTok{return}\NormalTok{ results}\OperatorTok{;}
\OperatorTok{\}}
\end{Highlighting}
\end{Shaded}

First, we create vectors for the output, which will serve as the columns
of the resulting DataFrame. Then, we iterate through the entries of
\texttt{data}. For each entry, we extract the summary statistics. We
calculate the standard deviation from the SST, round all statistics to
the nearest integer (with ties rounded to even), and convert them to
microseconds for better readability. Reporting three decimal digits
ensures the information remains precise to the nanosecond level. The
results are then added to the output vectors. Finally, we create the
\texttt{results} DataFrame using \texttt{DataFrame::create()}, setting
the class attributes to
\href{https://CRAN.R-project.org/package=rcpptimer}{\texorpdfstring%
{{\normalfont\fontseries{b}\selectfont rcpptimer}}%
{rcpptimer}} and \texttt{data.frame}, and settings the tags as row
names.

If \texttt{autoreturn} is \texttt{true} (the default), we write
\texttt{results} to R's global environment using the user provided
\texttt{name} (defaults to \texttt{times}). This implementation of
\texttt{stop()} makes it very convenient to use
\href{https://CRAN.R-project.org/package=rcpptimer}{\texorpdfstring%
{{\normalfont\fontseries{b}\selectfont rcpptimer}}%
{rcpptimer}}, as users are not required to handle the results themselves
(e.g.~by integrating them into their return objects). The timings will
be are available in R without adjusting the return statements of the
code that is to be timed.

In standard settings, users do not need to manually call
\texttt{stop()}. The destructor of the \texttt{Timer} class
automatically invokes \texttt{stop()} if \texttt{autoreturn} is set to
\texttt{true}, meaning the results are returned to R when the Timer
object goes out of scope (see the excerpt of ``rcpptimer.h'' in
\href{rcpptimer.h}{Section 3}). This is a very convenient feature, as
users do not need to worry about calling \texttt{stop()} manually.
Howevver, manually accessing the results is also possible. This is
possible as \texttt{aggregate()} and \texttt{stop()} return their
results. Furthermore, the \texttt{durations} and \texttt{tags} vectors
are public members of the \texttt{CppTimer} class, allowing users to
access the raw measurements if desired. This functionality is detailed
in the
\href{https://rcpptimer.berrisch.biz/articles/advanced.html\#accessing-unprocessed-data}{Advanced
Usage of rcpptimer} vignette.

In essence, users can simply utilize the \texttt{Timer} class by calling
\texttt{.tic()} and \texttt{.toc()}. The results are then automatically
returned to R, allowing users to obtain the results without writing a
single line of additional code.

\section{Other considerations}\label{other}

Beyond the core functionality described above, the development of
rcpptimer considers many other important aspects. Keeping the
performance of \texttt{.tic()} and \texttt{.toc()} is clearly among the
most important priorities. Other aspects include issuing warnings in
unexpected situations, enabling convenient scope-based timing and
providing advanced users with fine-grained control over the
functionality and contents of the Timer class. Several technical
details, such as minimizing lookups in the \texttt{tics} map, were also
carefully considered. The following subsections will explore a selection
of these aspects in greater detail.

\subsection{Warnings}\label{warnings}

Warnings are an important feature of this package, issued when the
\texttt{Timer} class is used in unintended ways. Four main cases are
considered:

\begin{itemize}
\tightlist
\item
  A \texttt{.toc()} call is not matched by a corresponding
  \texttt{.tic()} call.
\item
  A \texttt{.tic()} call is not matched by a corresponding
  \texttt{.toc()} call.
\item
  Multiple \texttt{.toc()} calls follow a single \texttt{.tic()} call.
\item
  Multiple \texttt{.tic()} calls occur without a matching
  \texttt{.toc()} call.
\end{itemize}

Minor adjustments to the \texttt{.toc()} implementation are sufficient
to allow the Timer class to identify the first three cases and issue
warnings. Catching the fourth situation would require a more complex
implementation. This is because multiple calls to \texttt{.tic()} must
be allowed to support repeated measurements of the same code block.
Identifying multiple \texttt{.tic()} calls without a \texttt{.toc()}
call in between would require some kind of interaction between
\texttt{.toc()} and \texttt{.tic()} calls and appropriate checks to
handle such events. We opted not to implement this feature, as it would
introduce unnecessary overhead in the \texttt{.tic()} method, which we
prioritize for speed. Moreover, the fourth situation is not particularly
problematic: if \texttt{.tic()} is called multiple times with the same
tag, the most recent call will be considered by the \texttt{.toc()}
method.

The first case, where a user places a \texttt{.toc()} call without a
matching \texttt{.tic()}, is the most crucial to handle. If this
situation were not caught, \texttt{.toc()} would try to access a
nonexistent entry in the \texttt{tics} map. In that case, the
\texttt{{[}{]}} operator creates an entry in that map and returns a
reference to this object. This would render the calculation of the
duration incorrect. The following code snipped demonstrates how we can
modify \texttt{.toc()} to catch this situation:

\begin{Shaded}
\begin{Highlighting}[]
\ControlFlowTok{if} \OperatorTok{(}\KeywordTok{auto}\NormalTok{ tic}\OperatorTok{\{}\NormalTok{tics}\OperatorTok{.}\NormalTok{find}\OperatorTok{(}\NormalTok{key}\OperatorTok{)\};}\NormalTok{ tic }\OperatorTok{!=}\NormalTok{ end}\OperatorTok{(}\NormalTok{tics}\OperatorTok{))}
\OperatorTok{\{}
\NormalTok{  nanoseconds duration }\OperatorTok{=}\NormalTok{ high\_resolution\_clock}\OperatorTok{::}\NormalTok{now}\OperatorTok{()} \OperatorTok{{-}}
                          \BuiltInTok{std::}\NormalTok{move}\OperatorTok{(}\NormalTok{tic}\OperatorTok{{-}\textgreater{}}\NormalTok{second}\OperatorTok{);}
\NormalTok{  durations}\OperatorTok{.}\NormalTok{push\_back}\OperatorTok{(}\NormalTok{duration}\OperatorTok{.}\NormalTok{count}\OperatorTok{());}
\NormalTok{  tags}\OperatorTok{.}\NormalTok{push\_back}\OperatorTok{(}\BuiltInTok{std::}\NormalTok{move}\OperatorTok{(}\NormalTok{key}\OperatorTok{.}\NormalTok{first}\OperatorTok{));}
\OperatorTok{\}}
\end{Highlighting}
\end{Shaded}

The core of the if statement is essentially the same as in the original
implementation. We added the if statement to check whether the key
exists in the \texttt{tics} map. If it does, we proceed as before. We
use an if statement with initializer here, a C++17 feature that allows
us declar and initialize variables directly in the if statement itself.
In this case, we declare \texttt{tic} and initialize it with the result
of the \texttt{std::map::find} method. \texttt{tic} is a
\texttt{std::map::iterator}, and we check if it is not equal to
\texttt{end(tics)}, i.e.~if \texttt{tic} points to an entry in our map.
If so, we can use \texttt{tic} to directly access the timepoint with
\texttt{tic-\textgreater{}second}. That is, we do not have to do a
second lookup, to access the timepopoint and calculate the duration,
making the process more efficient.

This approach is a modern and concise way to check if a key exists in a
map, and the if statement with initializer helps to keep the scope of
\texttt{tic} tight. This implementation handles the first case by simply
skipping the duration calculation. However, we want to handle this
situation more gracefully by issuing a warning. We can achieve this by
extending the if statement with an else block:

\begin{Shaded}
\begin{Highlighting}[]
\ControlFlowTok{else}
\OperatorTok{\{}
\NormalTok{  missing\_tics}\OperatorTok{.}\NormalTok{insert}\OperatorTok{(}\BuiltInTok{std::}\NormalTok{move}\OperatorTok{(}\NormalTok{key}\OperatorTok{.}\NormalTok{first}\OperatorTok{))}
\OperatorTok{\}}
\end{Highlighting}
\end{Shaded}

This writes the \texttt{tag} of the missing \texttt{.tic()} call into a
\texttt{std::set}. The set is a protected member of the
\texttt{CppTimer} class. We use a \texttt{std::set} for two reasons.
First, we can not use \texttt{Rcpp::warning} directly, as it is not
thread-safe and \texttt{.toc()} may be called in parallelized code
\citep{nagler2018r}. Second, the set only stores unique values. That is,
if a tag is added multiple times, it is only stored once. This is
intentional; we want to issue only one warning per tag.

Identifying \texttt{.tic()} calls which are not followed by
\texttt{.toc()} statement is straightforward. Initially, we used
\texttt{std::map::erase()} for this:

\begin{Shaded}
\begin{Highlighting}[]
\ControlFlowTok{if} \OperatorTok{(}\KeywordTok{auto}\NormalTok{ tic}\OperatorTok{\{}\NormalTok{tics}\OperatorTok{.}\NormalTok{find}\OperatorTok{(}\NormalTok{key}\OperatorTok{)\};}\NormalTok{ tic }\OperatorTok{!=}\NormalTok{ end}\OperatorTok{(}\NormalTok{tics}\OperatorTok{))}
\OperatorTok{\{}
\NormalTok{  nanoseconds duration }\OperatorTok{=}\NormalTok{ high\_resolution\_clock}\OperatorTok{::}\NormalTok{now}\OperatorTok{()} \OperatorTok{{-}}
                          \BuiltInTok{std::}\NormalTok{move}\OperatorTok{(}\NormalTok{tic}\OperatorTok{{-}\textgreater{}}\NormalTok{second}\OperatorTok{);}
\NormalTok{  durations}\OperatorTok{.}\NormalTok{push\_back}\OperatorTok{(}\NormalTok{duration}\OperatorTok{.}\NormalTok{count}\OperatorTok{());}
\NormalTok{  tics}\OperatorTok{.}\NormalTok{erase}\OperatorTok{(}\NormalTok{tag}\OperatorTok{);} \CommentTok{// Added }
\NormalTok{  tags}\OperatorTok{.}\NormalTok{push\_back}\OperatorTok{(}\BuiltInTok{std::}\NormalTok{move}\OperatorTok{(}\NormalTok{tag}\OperatorTok{));}
\OperatorTok{\}}
\end{Highlighting}
\end{Shaded}

So we removed the \texttt{tag} from the \texttt{tics} map. In
consequence, any remaining entries in the \texttt{tics} map referred to
unmatched \texttt{.tic()} calls. However, two problems arise from this
approach. First, calling \texttt{.toc()} multiple times (without calling
\texttt{.tic()} first) creates entries in \texttt{missing\_tics}. This
will produce warnings about missing \texttt{tic} statements, although
the real problem is that the user mistakenly placed multiple
\texttt{toc} statements in the same timer. Second, the \texttt{.erase()}
operation is costly, as it induces rebalancing of the map's internal
tree structure. Moreover, it is likely that subsequent calls to
\texttt{.tic()} with the same tag will follow (e.g.~if the timer is
placed in a loop), leading to repeated rebalancing. Both problems can be
solved by assigning a special value to the entry of the map instead of
using \texttt{erase()} to remove it. This way, we can identify unmatched
\texttt{.tic()} calls by inspecting the values of the map later while
avoiding the rebalancing operations in \texttt{.toc()} (and possibly
subsequent calls to \texttt{.tic()}). The following code snippet shows
the modified \texttt{.toc()} method:

\begin{Shaded}
\begin{Highlighting}[]
\NormalTok{durations}\OperatorTok{.}\NormalTok{push\_back}\OperatorTok{(}\NormalTok{duration}\OperatorTok{.}\NormalTok{count}\OperatorTok{());}
\NormalTok{tic}\OperatorTok{{-}\textgreater{}}\NormalTok{second }\OperatorTok{=}\NormalTok{ tic}\OperatorTok{{-}\textgreater{}}\NormalTok{second}\OperatorTok{.}\NormalTok{max}\OperatorTok{();} \CommentTok{// Replaces the erase operation}
\NormalTok{tags}\OperatorTok{.}\NormalTok{push\_back}\OperatorTok{(}\BuiltInTok{std::}\NormalTok{move}\OperatorTok{(}\NormalTok{tag}\OperatorTok{));}
\end{Highlighting}
\end{Shaded}

The code above uses the \texttt{tic} iterator to assign
\texttt{std::chrono::time\_point::max} to the entry of the map, which
represents the largest possible time duration\footnote{https://en.cppreference.com/w/cpp/chrono/time\_point/max}.
With this implementation, unmatched \texttt{.tic()} calls can be
identified by inspecting the values of the tics map. Entries with values
that differ from \texttt{std::chrono::time\_point::max} correspond to
\texttt{.tic()} calls that have not been matched by a \texttt{.toc()}
call. In these cases, we can issue a warning for the relevant tags. If
all timers are stopped correctly, the values of the tics map will be
\texttt{std::chrono::time\_point::max} for all tags.

Assigning the largest possible time point also solves the third case,
where we want to issue warnings for multiple \texttt{.toc()} calls
corresponding to a single \texttt{.tic()} call. As \texttt{.toc()} is
executed for the same tag again, the calculated duration will be
negative, as it results from subtracting the largest possible time point
(which is certainly in the future) from the current time. Since negative
entries in the \texttt{durations} vector already indicate the issue, we
do not have to modify the \texttt{.toc()} method further. Instead, we
adjusted \texttt{.aggregate()} to catch this situation. The following
code snippet shows the adjusted part of the \texttt{.aggregate()}
method:

\begin{Shaded}
\begin{Highlighting}[]
\ControlFlowTok{if} \OperatorTok{(}\NormalTok{duration }\OperatorTok{\textless{}} \DecValTok{0}\OperatorTok{)}
\OperatorTok{\{}
\NormalTok{  needless\_tocs}\OperatorTok{.}\NormalTok{insert}\OperatorTok{(}\NormalTok{tags}\OperatorTok{[}\NormalTok{i}\OperatorTok{]);}
  \ControlFlowTok{continue}\OperatorTok{;}
\OperatorTok{\}}
\end{Highlighting}
\end{Shaded}

For negative durations, we store the corresponding tag in another
\texttt{std::set}, which is also a protected member of the
\texttt{CppTimer} class. As mentioned above, using a \texttt{std::set}
ensures that the same warning is not issued multiple times for the same
tag.

The warnings are then handled by the \texttt{.print\_warnings()} method.
We will not present the implementation here to keep this paper concise,
but it simply iterates over the \texttt{missing\_tics} and
\texttt{needless\_tocs} sets, as well as the \texttt{tics} map, checking
for entries that are not equal to
\texttt{std::chrono::time\_point::min}. The \texttt{.print\_warnings()}
method is part of the \texttt{Timer} class and uses
\texttt{Rcpp::warning} to issue the warnings. It is automatically called
by the destructor of the \texttt{Timer} class if \texttt{verbose} is set
to \texttt{true}, so users do not need to call it manually. However,
they can disable the warnings by setting \texttt{verbose} to false.

\subsection{ScopedTimer}\label{ScopedTimer}

Timing scopes typically require placing a \texttt{.tic()} and
\texttt{.toc()} call at the beginning and end of the scope, which can be
cumbersome. To streamline this process, we implemented the
\texttt{ScopedTimer} class. This class is a subclass of
\texttt{CppTimer} and uses its own destructor to automatically call
\texttt{.toc()}. The following code snippet shows the implementation of
the subclass:

\begin{Shaded}
\begin{Highlighting}[]
\KeywordTok{class}\NormalTok{ ScopedTimer}
\OperatorTok{\{}
\KeywordTok{private}\OperatorTok{:}
\NormalTok{  CppTimer }\OperatorTok{\&}\NormalTok{timer}\OperatorTok{;}
\NormalTok{  string tag}\OperatorTok{;}

\KeywordTok{public}\OperatorTok{:}
\NormalTok{  ScopedTimer}\OperatorTok{(}\NormalTok{CppTimer }\OperatorTok{\&}\NormalTok{timer}\OperatorTok{,}\NormalTok{ string tag }\OperatorTok{=} \StringTok{"scoped"}\OperatorTok{)} \OperatorTok{:}\NormalTok{ timer}\OperatorTok{(}\NormalTok{timer}\OperatorTok{),}\NormalTok{ tag}\OperatorTok{(}\NormalTok{tag}\OperatorTok{)}
  \OperatorTok{\{}
\NormalTok{    timer}\OperatorTok{.}\NormalTok{tic}\OperatorTok{(}\NormalTok{string}\OperatorTok{(}\NormalTok{tag}\OperatorTok{));}
  \OperatorTok{\}}
  \OperatorTok{\textasciitilde{}}\NormalTok{ScopedTimer}\OperatorTok{()}
  \OperatorTok{\{}
\NormalTok{    timer}\OperatorTok{.}\NormalTok{toc}\OperatorTok{(}\NormalTok{string}\OperatorTok{(}\NormalTok{tag}\OperatorTok{));}
  \OperatorTok{\}}
\OperatorTok{\};}
\end{Highlighting}
\end{Shaded}

The \texttt{ScopedTimer} class features a single constructor that
initializes its two private members. The first member is a reference to
a \texttt{CppTimer} object, and the second is a \texttt{std::string}
that serves as the tag for the timer (defaulting to ``scoped''). Upon
construction, the constructor calls \texttt{.tic()} on the provided
\texttt{CppTimer} instance, while \texttt{.toc()} is automatically
invoked when the \texttt{ScopedTimer} instance is destroyed (i.e.~when
it goes out of scope). This behaviour is crucial, as the destructor's
call to \texttt{.toc()} ensures that the timer is stopped automatically.
This is very convenient, as users do not have to worry about calling
\texttt{.toc()} manually.

\subsection{Resetting Summary Statistics}\label{reset}

As described in \hyperref[SummaryStatistics]{Section 3.3}, the
implementation of the \texttt{aggregate()} method makes it possible to
update the summary statistics as new durations get observed. However, in
some scenarios, it may be useful to reset these summary statistics. This
can be achieved by calling the reset() method, which clears the stored
data and allows users to start fresh. Here is how the \texttt{.reset()}
method is implemented:

\begin{Shaded}
\begin{Highlighting}[]
\DataTypeTok{void}\NormalTok{ reset}\OperatorTok{()}
\OperatorTok{\{}
\NormalTok{  tics}\OperatorTok{.}\NormalTok{clear}\OperatorTok{(),}\NormalTok{ durations}\OperatorTok{.}\NormalTok{clear}\OperatorTok{(),}\NormalTok{ tags}\OperatorTok{.}\NormalTok{clear}\OperatorTok{(),}\NormalTok{ data}\OperatorTok{.}\NormalTok{clear}\OperatorTok{();}
\OperatorTok{\}}
\end{Highlighting}
\end{Shaded}

The implementation is straightforward. We just clear the \texttt{tics},
\texttt{durations}, \texttt{tags} and \texttt{data} objects. This is
sufficient to remove all timings from the \texttt{CppTimer} class. The
\texttt{reset()} method is part of the \texttt{CppTimer} class.

\subsection{Fragmentation of the Package}\label{fragmentation}

As shown in \hyperref[implementation]{Section 3}, we fragmented the code
of this project into two repositories. This project originated from the
\href{https://CRAN.R-project.org/package=RcppClock}{\texorpdfstring%
{{\normalfont\fontseries{b}\selectfont RcppClock}}%
{RcppClock}} R package, where all the C++ code was contained in a single
header file. However, the core timing functionality did not depend on
\href{https://CRAN.R-project.org/package=Rcpp}{\texorpdfstring%
{{\normalfont\fontseries{b}\selectfont Rcpp}}%
{Rcpp}}. Therefore, we decided to split the code into two parts. The
core functionality of the timer was moved to a separate repository,
removing any Rcpp dependencies and leaving a standalone header-only
library that provides the \texttt{CppTimer} class, which contains all
R-agnostic parts. This project and the corresponding GitHub repository
are called \texttt{cpptimer}.

The \href{https://CRAN.R-project.org/package=rcpptimer}{\texorpdfstring%
{{\normalfont\fontseries{b}\selectfont rcpptimer}}%
{rcpptimer}} package contains the R-specific parts. It contains the
\texttt{Timer} class, which inherits from \texttt{CppTimer}. We
intentionally added the \texttt{Timer} class to the
\href{https://CRAN.R-project.org/package=Rcpp}{\texorpdfstring%
{{\normalfont\fontseries{b}\selectfont Rcpp}}%
{Rcpp}} namespace so that instances of this class can be constructed
using \texttt{Rcpp::Timer}, which aligns closely with the package name.
The \texttt{Timer} class contains constructors, the \texttt{.stop()} and
\texttt{.print\_warnings()} methods, and the destructor.

This fragmentation makes it easy to write interfaces to
\texttt{CppTimer} for other programming languages. Additionally, when
implementing an interface for another programming language, the
\href{https://CRAN.R-project.org/package=rcpptimer}{\texorpdfstring%
{{\normalfont\fontseries{b}\selectfont rcpptimer}}%
{rcpptimer}} R package can serve as a reference implementation for such
interfaces.

\section{Conclusion}\label{conclusion}

The rcpptimer package offers a versatile and efficient solution for
timing (R)Cpp code within the R ecosystem. The package fills a critical
gap left by existing timing tools by integrating support for OpenMP,
nested, and overlapping timers. The automatic return of aggregated
timing results to R further streamlines the benchmarking process,
reducing the need for custom code. Its simplicity makes it accessible to
users who are not proficient in C++, as there is no need to manually
handle the results, calculate summary statistics, issue warnings or set
up dependencies. Furthermore, its advanced features ensure its
usefulness in more complex situations.

The package is designed to work in different settings which involve
\href{https://CRAN.R-project.org/package=Rcpp}{\texorpdfstring%
{{\normalfont\fontseries{b}\selectfont Rcpp}}%
{Rcpp}}. It can be used with \texttt{Rcpp::cppFunction}, with
\texttt{Rcpp::sourceCpp}, and with R packages. It is easy to use, as
users only need to call \texttt{.tic()} and \texttt{.toc()} to time
their code. The results are then automatically returned to R. The
package is also flexible, allowing fine-grained control over the return
objects. For example, users can call \texttt{aggregate()} and
\texttt{stop()} themselves if they want to handle the results manually.
The package is also very efficient, as its design prioritizes minimized
overhead. This is achieved, in particular, by keeping the
\texttt{.tic()} and \texttt{.toc()} methods as simple as possible.

The development of
\href{https://CRAN.R-project.org/package=rcpptimer}{\texorpdfstring%
{{\normalfont\fontseries{b}\selectfont rcpptimer}}%
{rcpptimer}} follows the best practices of R package development. We
test our package using
\href{https://CRAN.R-project.org/package=testthat}{\texorpdfstring%
{{\normalfont\fontseries{b}\selectfont testthat}}%
{testthat}} to ensure that the package works as intended. We test that
the summary statistics are correctly calculated, that the warnings align
with the code, that \texttt{print.rcpptimer} scales as intended, that
\texttt{aggregate()} correctly updates the results when called
subsequently as new durations emerge, and many more. The tests have a
coverage of 100\% (evaluated using
\href{https://CRAN.R-project.org/package=covr}{\texorpdfstring%
{{\normalfont\fontseries{b}\selectfont covr}}%
{covr}}). Furthermore, we use GitHub Actions to test the package
continuously on multiple platforms.

This paper focused on the package's \emph{implementation} and
\emph{design}; therefore, the \emph{usage} of the package was only
discussed a little. However, we provide extensive documentation on the
usage in multiple vignettes. This includes a detailed introduction to
the package, which covers the usage of \texttt{Rcpp::Timer} with
\texttt{Rcpp::CppFunction}, setting up multiple, nested, and overlapping
timers, and using the \texttt{ScopedTimer} class\footnote{https://rcpptimer.berrisch.biz/articles/rcpptimer.html}.
Furthermore, we provide information on how to use the package with
\texttt{Rcpp::sourceCpp}\footnote{https://rcpptimer.berrisch.biz/articles/sourceCpp.html},
we discuss how
\href{https://CRAN.R-project.org/package=rcpptimer}{\texorpdfstring%
{{\normalfont\fontseries{b}\selectfont rcpptimer}}%
{rcpptimer}} can this to your own package and what to consider when
doing so\footnote{https://rcpptimer.berrisch.biz/articles/packages.html},
we show how manual access to the results is possible and how the
autoreturn feature can be disabled\footnote{https://rcpptimer.berrisch.biz/articles/autoreturn.html},
and, finally, we discuss advanced topics like accessing the unprocessed
timings, updating the timer and resetting the timer\footnote{https://rcpptimer.berrisch.biz/articles/advanced.html}.

Compared to alternatives like
\href{https://CRAN.R-project.org/package=tictoc}{\texorpdfstring%
{{\normalfont\fontseries{b}\selectfont tictoc}}%
{tictoc}},
\href{https://CRAN.R-project.org/package=microbenchmark}{\texorpdfstring%
{{\normalfont\fontseries{b}\selectfont microbenchmark}}%
{microbenchmark}}, and
\href{https://CRAN.R-project.org/package=bench}{\texorpdfstring%
{{\normalfont\fontseries{b}\selectfont bench}}%
{bench}},
\href{https://CRAN.R-project.org/package=rcpptimer}{\texorpdfstring%
{{\normalfont\fontseries{b}\selectfont rcpptimer}}%
{rcpptimer}} provides insights into C++ code, including parallelized
tasks. That is, it is suitable for timing execution times of C++ code
sections without considering the overhead of calling the functions from
R. It bridges the gap between R and C++ benchmarking tools, offering
precision and flexibility. This positions
\href{https://CRAN.R-project.org/package=rcpptimer}{\texorpdfstring%
{{\normalfont\fontseries{b}\selectfont rcpptimer}}%
{rcpptimer}} as a valuable asset for developers seeking to optimize
performance in
\href{https://CRAN.R-project.org/package=Rcpp}{\texorpdfstring%
{{\normalfont\fontseries{b}\selectfont Rcpp}}%
{Rcpp}} code.

\bibliography{RJreferences.bib}

\end{document}